\documentclass[aps,pra,groupaddress,showpacs,twocolumn]{revtex4-1}
\usepackage{graphicx,bm,amsmath,amsfonts,bbold, amssymb,color,microtype}

\newcommand{\be}{\begin{equation}}
\newcommand{\ee}{\end{equation}}
\newcommand{\bea}{\begin{eqnarray}}
\newcommand{\eea}{\end{eqnarray}}
\newcommand{\im}[1]{\mbox{Im}[#1]}

\begin{document}

\title{Contrasting Spectral Signatures and Sensitivities of CPA-Lasing \\in a $\cal PT$-Symmetric Periodic Structure}

\author{Li Ge}
\email{li.ge@csi.cuny.edu}
\affiliation{\textls[-18]{Department of Engineering Science and Physics, College of Staten Island, CUNY, Staten Island, NY 10314, USA}}
\affiliation{The Graduate Center, CUNY, New York, NY 10016, USA}
\author{Liang Feng}
\affiliation{Department of Electrical Engineering, The State University of New York at Buffalo, Buffalo, NY 14260, USA}

\begin{abstract}
The CPA-laser is a coexisting state of coherent perfect absorption and lasing that was proposed in parity-time ($\cal PT$) symmetric photonic systems. In this work we show that the spectral signature of a CPA-laser displayed by the singular value spectrum of the scattering matrix ($S$) can be orders of magnitude wider than that displayed by the eigenvalue spectrum of $S$. Since the former reflects how strongly light can be absorbed or amplified and the latter announces the spontaneous symmetry breaking of $S$, these contrasting spectral signatures indicate that near perfect absorption and extremely strong amplification can be achieved even in the $\cal PT$-symmetric phase of $S$, which is known for and defined by its flux-conserving eigenstates. We also show that these contrasting spectral signatures are accompanied by strikingly different sensitivities to disorder and imperfection, suggesting that the eigenvalue spectrum is potentially suitable for sensing and the singular value spectrum for robust switching.
\end{abstract}


\maketitle

\section{Introduction}

The study of parity-time ($\cal{PT}$) symmetric systems \cite{Bender1,Bender2,Bender3} has become one of the fastest growing fields in photonics during the past few years. While a true $\cal PT$-symmetric photonic system requires judiciously balanced gain and loss elements such that its refractive index satisfies $n(x)=n^*(-x)$ \cite{El-Ganainy_OL06,Moiseyev,Kottos,Musslimani_prl08,Makris_prl08,Regensburger,Lin,conservation,Feng2,Robin,PT2D}, many of its novel properties stem from the underlying exceptional points \cite{EP1,EP2,EPMVB,EP3,EP4,EP5,EP6,EP7,EP8} where two or more non-Hermitian eigenstates and their eigenvalues coalesce. As a result, the demonstrations of $\cal PT$-symmetric systems often utilized non-balanced gain and loss, including experiments on loss-enhanced transmission \cite{EP7} and $\cal PT$-symmetric lasers \cite{EP_CMT,PTlaser_nonlinear,Feng,Hodaei}. This approach removes the stringent requirement on fine tuning the imaginary part of the refractive index, which also has the benefit of using loss-dominant structures to reduce the noise coming from amplified spontaneous emission.

Nevertheless, there are unique properties that are conveniently found in true $\cal PT$-symmetric systems but not in their counterparts with unbalanced gain and loss. One of them is the generalized conservation law for wave propagation in one-dimensional (1D) and quasi-1D systems \cite{conservation,conservation2D}, which was demonstrated in a $\cal PT$-symmetric RLC circuit \cite{circuit}. Another one is the existence of CPA-laser points \cite{Longhi}. CPA, or coherent perfect absorption \cite{CPA,CPA_exp}, utilizes coherent illumination to suppress transmission and reflection, such that light is completely trapped in a pre-determined spatial domain to achieve perfect absorption. While based on the same principle as critical coupling (i.e., destructive interference) \cite{cc}, CPA is a more general principle that can be viewed as a time-reversed lasing process and applies to arbitrarily complex systems, even if they are strongly scattering. If such a system is $\cal PT$-symmetric, then it was predicted that CPA must occur at a lasing frequency of the system \cite{CPALaser}, which is referred to as a CPA-laser. This intriguing phenomenon was envisioned to provide the ultimate extinction ratio for the ``on" and ``off" states of coherent signals, which was successfully demonstrated in a recent experiment \cite{CPALaser_exp}.
%which can be conveniently achieved by switching between the lasing state and the CPA state via controlling the relative phase of two incoming beams.

In this work we show that a CPA-laser can display contrasting spectral signatures when probed using different methods. Since a CPA-laser exists only in a $\cal PT$-broken phase of the eigenvalue spectrum of the scattering matrix ($S$) \cite{CPALaser}, one can refer to the width $W$ of this $\cal PT$-broken phase as a spectral signature of a CPA-laser. This approach requires locating the wavelengths of the two exceptional points that delineate this $\cal PT$-broken phase. Alternatively, one can also probe a CPA-laser by observing the minimal and maximal output intensity at different wavelengths, which can be tuned into by changing the relative phase of the coherent illumination beams \cite{Longhi}. Such intensity information is contained by the singular value spectrum of $S$, which reflects directly how strong absorption and amplification are. As we shall see, when the singular values of $S$ become unimodular simultaneously, the output intensity does not depend on the relative phase or amplitude of the input beams, where amplification and absorption disappear. The spectral signature of a CPA-laser can hence also be quantified by the wavelength difference $V$ between two such input-insensitive points that confine a CPA-laser point.

It is striking that these two measures, $W$ and $V$, can differ by several orders of magnitude, which indicates that near perfect absorption and extremely strong amplification can be achieved even in the $\cal PT$-symmetric phase of the scattering matrix, which is known for and defined by its flux-conserving eigenstates.

We also show that these contrasting spectral signatures of a CPA-laser are accompanied by very different sensitivities to imperfection. With a minute change of the imaginary part of the refractive index from the CPA-laser condition, the $\cal PT$-broken phase can be completely eliminated, while the singular value spectrum of $S$ continues to show near perfect absorption and extremely strong amplification. In addition, the eigenstates of $S$ in the $\cal PT$-broken phase lead to a strong sensitivity to the relative amplitude of the input beams, whereas the singular vectors of $S$ give rise to a strong sensitivity to the relative phase of the input beams. Even though $\cal PT$ symmetry is destroyed in the presence of disorder, we show that the singular value spectrum maintains an impressive intensity contrast between the amplifying and absorbing states. Therefore, one can potentially utilize the eigenvalue spectrum of $S$ near a CPA-laser point for sensing and its singular value spectrum for robust switching.

%Another issue is how to switch between CPA and laser: the corresponding eigenstates of the scattering matrix feature the same relative phase $\pi/2$ between the two incoming beams. Therefore, the switching between CPA and laser cannot be achieved by tuning the relative phase of the two incoming beams; a fine tuning of their relative amplitude is needed instead.

Below we illustrate these results in a 1D $\cal PT$-symmetric periodic structure (see Fig.~\ref{fig:schematic}), which was used in the CPA-laser experiment mentioned above \cite{CPALaser_exp}. Using an approximate form of $S$ derived from a coupled mode theory, we first show qualitatively that the CPA-laser in this system appears as a singular feature in the eigenvalue spectrum of $S$, i.e., the aforementioned measure $W$ vanishes. In the meanwhile, the spectral width $V$ displayed by the singular value spectrum is much more extended. We then verify these findings by deriving an exact form of $S$, which reveals that $W$ and $V$ follow different scaling behaviors with the system size, leading to their vastly different values in a large system. Finally we study the sensitivities of the CPA-laser using this exact form of $S$, and we comment on the general properties of the singular values of $S$ in a $\cal PT$-symmetric system. In the appendix we illustrate the forming of the CPA-laser points at the boundary of the Brillouin zone using the motions of the poles and zeros of $S$.

\section{Spectral signatures of a CPA-laser}

\begin{figure}[b]
\centering
\includegraphics[width=\linewidth]{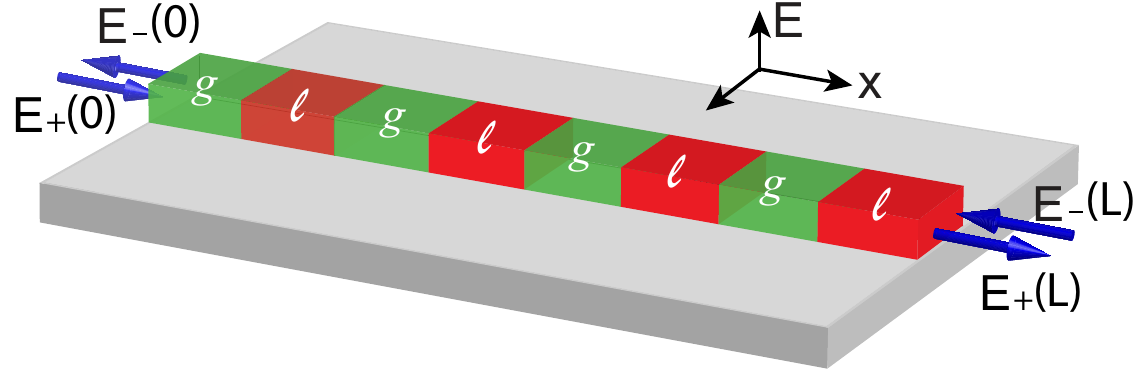}
\caption{(Color online) Schematic of a 1D $\cal PT$-symmetric periodic structure, consisting of alternate layers of gain (\textit{g}) and loss (\textit{l}) regions.} \label{fig:schematic}
\end{figure}

We resort to both transfer matrix and scattering matrix methods in our analyses below. As we will show shortly, while these two methods are equivalent to each other, they offer different advantages in the study of a CPA-laser. More specifically, the singular value and eigenvalue spectra of the scattering matrix illustrate the spectral signatures of a CPA-laser, whereas the transfer matrix method provides a convenient way to derive the relation between incident and scattered light amplitudes and hence $S$ as well.

We base our discussion on the scalar Helmholtz equation of the electric field
\begin{equation}
  \left[\nabla^2 + n^2(x)\, k_0^2\right] E(x;k_0) = 0,\label{eq:Helmholtz}
\end{equation}
which describes steady-state solutions for transverse electric waves in 1D systems. Here $k_0$ is the free-space wave number and $n(x)$ is the refractive index of the periodic structure. We consider a constant refractive index $n_j\equiv n_r+in_i\,(j=1,3,5\ldots)$ in each loss layer and a constant refractive index $n_{j+1}$ in each gain layer (see Fig.~\ref{fig:schematic}), and $\cal PT$ symmetry is satisfied by requiring $n_{j}=n_{j+1}^*$. The length of each layer is $a/2$ where $a$ is the lattice constant, and the whole structure has $N$ layers with a total length of $L=Na$. Outside the $\cal PT$-symmetric structure we assume the refractive index is $n_0$.

\subsection{Coupled mode theory}
\label{sec:CMT}
We assume that the $\cal PT$-symmetric periodic system is differentiated from the leads only by the weak modulation of the imaginary part of its refractive index, i.e., $n_r=n_0\gg n_i$. Under this condition, we can take the following ansatz
\be
E(x;k_0) = E_+(x)\exp(ikx) + E_-(x)\exp(-ikx),\label{eq:E}
\ee
where $k=n_rk_0$ is the effective wave number in the system and $E_\pm(x)$ are the slowing-varying envelopes (SVE) of the electrical fields propagating in the forward and backward directions. To derive the coupling between $E_+$ and $E_-$, we express the modulation of the refractive index by the summation of its Fourier components, i.e., $\im{n(x)}=\Sigma_{m}f_m\exp(imqx)$, where $m$ is an integer from $-\infty$ to $\infty$ and $q\equiv2\pi/a$ is the lattice wave number. We note that $f_0=0$ due to $\cal PT$-symmetry, and near the boundary of the Brillouin zone, i.e., $k\approx q/2$, the scattering between the forward and backward propagating waves is mainly due to the $m=\pm 1$ components of $\im{n(x)}$.

By keeping only these first-order scattering processes, we substitute $E(x;k_0)$ in Eq.~(\ref{eq:Helmholtz}) by the ansatz~(\ref{eq:E}) and perform the SVE approximation, neglecting terms containing $d^2E_\pm/dx^2$ as well as $n_i^2$. The result is the following coupled mode equation \cite{CPALaser_exp}:
\begin{align}
i\frac{d}{dx}
\begin{pmatrix}
E_- \\
E_+
\end{pmatrix}
=
\begin{pmatrix}
0 & -\gamma\\
-\gamma & 0
\end{pmatrix}
\begin{pmatrix}
E_- \\
E_+
\end{pmatrix}
\equiv H
\begin{pmatrix}
E_- \\
E_+
\end{pmatrix}
.\label{eq:H}
\end{align}
Here $H$ is the effective Hamiltonian of the system, and $\gamma=2n_ik_0/\pi$ is the coupling between $E_\pm$. By applying the evolution operation $\exp(-iHx)$, the transfer matrix $M$, defined by
\be
\begin{pmatrix}
E_- \\
E_+
\end{pmatrix}_{x=0}
=
M
\begin{pmatrix}
E_- \\
E_+
\end{pmatrix}_{x=L},
\ee
is simply
\be
M=
\begin{pmatrix}
\cos\gamma L & i\sin\gamma L\\
i\sin\gamma L & \cos\gamma L
\end{pmatrix}
\equiv
\begin{pmatrix}
M_{11} & M_{12}\\
M_{21} & M_{22}
\end{pmatrix}.
\ee
A lasing mode features an outgoing boundary condition, i.e.,
$E_+(0),E_-(L)=0$, which requires $M_{22}=0$; a CPA state features an incoming boundary condition, i.e.,
$E_+(L),E_-(0)=0$, which requires $M_{11}=0$. As can be seen from the expression above, $M_{11}$ and $M_{22}$ are identical and they vanish simultaneously in this $\cal PT$-symmetric periodic structure when $\cos\gamma L=0$. This condition is satisfied at the boundary of the Brillouin zone for discrete values of the gain and loss strength
\be
n_i = \frac{n_r\pi}{4N}(2P-1)\quad(P=1,2,\ldots),\label{eq:n_i}
\ee
which does not depend on the value of the lattice constant $a$. In the appendix we show that these values of $n_i$ are due to two different scenarios of pole (and zero) motions of the scattering matrix.

Next we discuss the spectral signature of this CPA-laser point displayed by the eigenvalue spectrum of the scattering matrix, defined by
\be
|\Phi\rangle
=
S
|\Psi\rangle, \,\,
|\Phi\rangle
=\begin{bmatrix}
E_-(0) \\
E_+(L)
\end{bmatrix},\,\,
|\Psi\rangle
=
\begin{bmatrix}
E_+(0) \\
E_-(L)
\end{bmatrix}.
\ee
$S$ can exhibit spontaneous $\cal PT$ symmetry breaking \cite{CPALaser}: In the $\cal PT$-symmetric phase, its two eigenstates are both flux conserving, i.e., the corresponding eigenvalues $s_\pm(k_0)$ satisfy $|s_\pm(k_0)|=1$; in the $\cal PT$-broken phase where a CPA-laser exists, they satisfy $|s_+(k_0)|=|s_-(k_0)|^{-1}>1$ instead. The lasing state occurs when $|s_+(k_0)|\rightarrow \infty$, and at the same time the CPA state can be realized in the eigenstate corresponding to $|s_-(k_0)|\rightarrow 0$. In a multi-layer system, there can be multiple regions of the $\cal PT$-broken phase \cite{CPALaser}, and each CPA-laser point is confined by such a $\cal PT$-broken phase. Here we focus on the one at the boundary of the Brillouin zone and refer to its width $W$ (in terms of the free-space wavelength $\lambda_0=2\pi/k_0$) as one spectral signature of the CPA-laser in our system.

\begin{figure}[b]
\centering
\includegraphics[width=\linewidth]{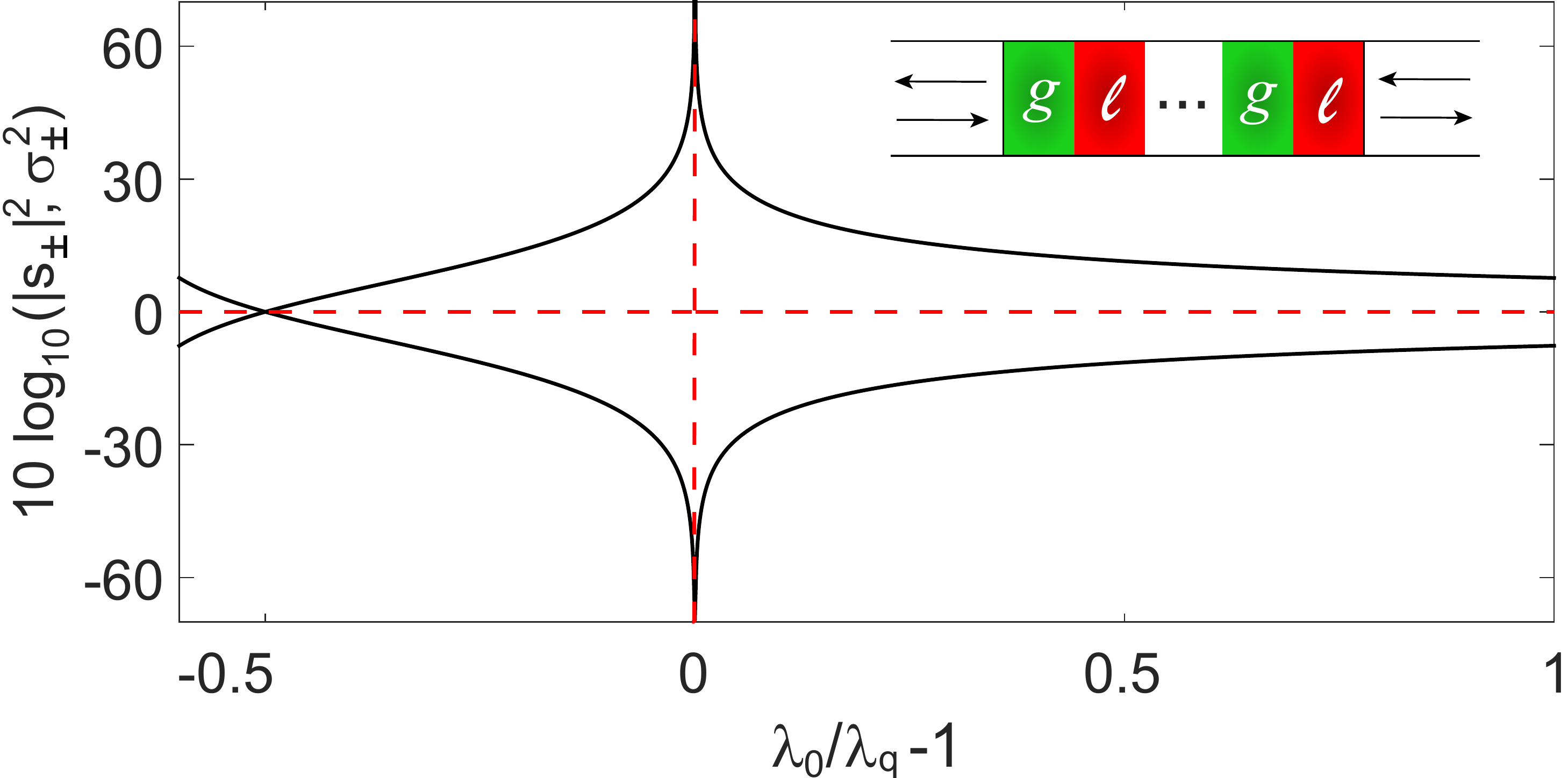}
\caption{(Color online) Eigenvalues (dashed lines) and singular values (solid lines) of the scattering matrix calculated using the coupled mode equation (\ref{eq:H}). %They are plotted as a function of the free-space wavelength $\lambda_0$ of the coherent input beams.
Inset: $\cal PT$-symmetric periodic structure similar to Fig.~\ref{fig:schematic}, consisting $1000$ alternate layers of gain and loss with refractive index $n_r=3.203\pm in_i$. $n_i$ is given by Eq.~(\ref{eq:n_i}) with $P=1$ and approximately $2.516\times10^{-3}$. $\lambda_q=2n_ra$ is the value of $\lambda_0$ at the CPA-laser point. The vertical section of the dashed line indicates the singular feature of the CPA-laser in the eigenvalue spectrum of the scattering matrix. The results shown are independent of the value of $a$.} \label{fig:CMT}
\end{figure}

The scattering matrix can be conveniently expressed using the elements of the transfer matrix, and in our case it is given by
\be
S
=
\frac{1}{\cos\gamma L}
\begin{pmatrix}
i\sin\gamma L & 1\\
1 & -i\sin\gamma L
\end{pmatrix}.\label{eq:S}
\ee
We note that this expression is valid everywhere except exactly at the CPA-laser point, where the denominator diverges. It is straightforward to see that the two eigenvalues of $S$ in Eq.~(\ref{eq:S}) are always unimodular: $s_\pm = {\pm\sqrt{1-\sin^2\gamma L}}/{\cos\gamma L} = \pm1$,
which means that the system is in the $\cal PT$-symmetric phase except right at the CPA-laser point. Therefore, we conclude that within the coupled mode theory, the CPA-laser at the boundary of the Brillouin zone in our $\cal PT$-symmetric periodic structure is a singular feature with a vanishing $W$ (see Fig.~\ref{fig:CMT}).

The spectral signature of the same CPA-laser point displayed by the singular value spectrum of $S$ is completely different. The singular values $\sigma_\pm$ and the left and right singular vectors ($\phi_\pm,\psi_\pm$) of $S$ are defined by
\begin{align}
S|\psi_\pm\rangle = \sigma_\pm|\phi_\pm\rangle, \quad S^\dagger|\phi_\pm\rangle = \sigma_\pm|\psi_\pm\rangle.\label{eq:svd}
\end{align}
where ``$\dagger$" denotes the Hermitian conjugate as usual. $\sigma_\pm$ are non-negative real quantities, and $\phi_\pm,\psi_\pm$ each form an orthogonal basis, satisfying $\phi_m^\dagger\phi_n=\psi_m^\dagger\psi_n=\delta_{mn}$. To illustrate the physical meanings of the singular value spectrum, we express the input amplitudes in the basis of $|\psi_\pm\rangle$ (i.e., $|\Psi\rangle = C_+|\psi_+\rangle+C_-|\psi_-\rangle$) and  consider the extinction ratio between the output and the input intensities, i.e., $\Theta = (|E_-(0)|^2+|E_+(L)|^2)/(|E_+(0)|^2+|E_-(L)|^2)$ \cite{Longhi}. It is easy to see that
\be
\Theta = \frac{\langle\Phi|\Phi\rangle}{\langle\Psi|\Psi\rangle} = \frac{\langle\Psi|S^\dagger S|\Psi\rangle}{\langle\Psi|\Psi\rangle} = \frac{\Sigma_{n=\pm}|C_n|^2\sigma_n^2}{\Sigma_{n=\pm}|C_n|^2}. \label{eq:ratio}
\ee
Therefore, the absorbing state is given by the right singular vector $|\psi_-\rangle$ with $\sigma_-^2<1$, and coherence perfect absorption takes place when $\sigma_-=0$. Similarly, the amplifying state is given by $|\psi_+\rangle$ with $\sigma_+^2>1$, and lasing is achieved when $\sigma^2_+\rightarrow\infty$.

We note that $\sigma_\pm$ satisfy $|\sigma_+\sigma_-|=1$ due to the $\cal PT$ symmetry of the system. This result can be derived in the following way. A $\cal PT$-symmetric $S$ satisfies $PS^*P = S^{-1}$ \cite{CPALaser}, where $P$ is the matrix form of the parity operator and given by the first Pauli matrix here. Now for a 1D system with reciprocity \cite{reciprocity}, $S$ is symmetric and hence $S^\dagger=S^*$. Therefore, by multiplying $P$ from the left to the second relation in Eq.~(\ref{eq:svd}) and inserting an identity matrix $\mathbb{1}=P^2$ after $S^\dagger$, we find
\be
S^{-1}P|\phi_\pm\rangle=\sigma_\pm|\psi_\pm\rangle.
\ee
This relation indicates two possibilities: either the left singular vectors $|\phi_\pm\rangle$ are parity eigenstates with $\sigma_\pm=\pm 1$, or $P|\phi_+\rangle = e^{i\theta}|\phi_-\rangle$ with $\sigma_+\sigma_-=e^{i\theta}$, both of which lead to $|\sigma_+\sigma_-|=1$. To verify this property directly, we note that the coupled mode theory gives $\sigma_\pm^2 = {(1\pm\sin{\gamma L})^2}/{\cos^2{\gamma L}}$ away from the CPA-laser point using Eq.~(\ref{eq:S}), which satisfy this property.

A consequence of $|\sigma_+\sigma_-|=1$ is that there are points where $\sigma_\pm$ become unimodular simultaneously, which takes place in the coupled mode theory when $\sin{\gamma L}=0$. At these places the extinction ratio $\Theta$ also becomes 1 and independent of $C_\pm$ as can be seen from Eq.~(\ref{eq:ratio}), which means that the output intensity is independent of the relative phase \textit{and} amplitude of the two input beams. This observation is a more general statement than what was pointed out in Ref.~\cite{CPA_exp}, where such points were found to be independent of the relative phase of the input beams. This general property is not unique to $\cal PT$-symmetric systems, although the fact that they always occur at $|\sigma_\pm|=1$ is. Another implication is that $|s_\pm|$ also go through these input-insensitive points, although in the present case it is not particular interesting since $|s_\pm|=1$ in the $\cal PT$-symmetric phase (see Fig.~\ref{fig:CMT}), where these input-insensitive points take place. Also note that the inverse of the statement is not true: $|\sigma_\pm|$ do not necessarily go through the points where $|s_\pm|$ become the same, whether or not the system is $\cal PT$-symmetric.

We then define another spectral signature, $V$, by the distance (in terms of $\lambda_0$) between two input-insensitive points that confine a CPA-laser point. In the coupled mode theory $V$ is infinite, because $|\sigma_\pm|$ becomes 1 only on the shorter wavelength side of the CPA-laser (see Fig.~\ref{fig:CMT}), i.e., at $\lambda_0=n_ra/P$ $(P=1,2,\ldots)$. Although the vanishing value of $W$ and the diverging value of $V$ are underestimated and overestimated in the coupled mode theory, we show below, by calculating the scattering matrix rigorously, that indeed the spectral signatures displayed by its eigenvalue spectrum and singular value spectrum are vastly different.

\subsection{Exact results}

To verify the qualitative knowledge gained from the coupled mode theory, we start in this section by deriving an exact expression of the transfer matrix $M$:
\be
M(k_0) = D_0^{-1} \left[\Pi_{j=1}^{2N} m_j\right] D_0.\label{eq:M}
\ee
$m_j$ are the transfer matrices of each layer, and again there are $N$ periods of gain and loss elements and a total of $2N$ layers. $D_0$ is given by
$(\begin{smallmatrix}1 & 1 \\n_0 & -n_0\end{smallmatrix})$ and $D_0^{-1}$ is its matrix inverse.
We note that a phase convention of $E_\pm(z)$ different from that in the coupled mode theory has been used in Eq.~(\ref{eq:M}), i.e., $E_\pm(L)\rightarrow\exp(\mp ikL)E_\pm(L)$, which however does not bear any physical significance.

With a constant refractive index in each layer, $m_j$ can be easily found to be:
\be
m_j(k_0) =
\begin{bmatrix}
\cos(n_jk_0a/2) & \frac{i}{n_j}\sin(n_jk_0a/2) \\
i n_j\sin(n_jk_0a/2) & \cos(n_jk_0a/2)
\end{bmatrix}.\nonumber
\ee
Denoting $s\equiv\sin(n_jk_0a/2), c\equiv\cos(n_jk_0a/2)$, the transfer matrix for one period is then given by
\be
M_j \equiv m_jm_{j+1} =
\begin{pmatrix}
|c|^2 - \frac{n_j^*}{n_j}|s|^2 & i\frac{sc^*}{n_j}-c.c. \\
in_jsc^*-c.c. & |c|^2 - \frac{n_j}{n_j^*}|s|^2
\end{pmatrix},\label{eq:M_period}
\ee
where ``$c.c.$" stands for complex conjugate. It is easy to check that $\det(M_j)=1$ and it satisfies the constraints imposed by $\cal PT$ symmetry \cite{Longhi}:
\be
m_{11}(k_0) = m_{22}^*(k_0),\quad m_{12,21}(k_0) = -m_{12,21}^*(k_0),\label{eq:symm}
\ee
where $m_{pq}\,(p,q=1,2)$ are the four elements of $M_j$. Note that $k_0$ is real here, and Eq.~(\ref{eq:M_period}) needs to be modified if a complex $k_0$ is considered, for example, in the calculation of the poles and zeros of the $S$ matrix (see the Appendix). For a unimodular matrix, its $N$th power can be calculated analytically \cite{BornWolf}:
\be
M_j^N =
\begin{pmatrix}
m_{11}U_{N-1}-U_{N-2} & m_{12}U_{N-1} \\
m_{21} U_{N-1} & m_{22}U_{N-1}-U_{N-2}
\end{pmatrix},\label{eq:iterative}
\ee
and here $U_N$ are the Chebyshev polynomials of the second kind with argument
%$\theta=\cos^{-1}[(m_{11}+m_{22})/2]$:
%\be
%U_N(\theta) = \frac{\sin[(N+1)\theta]}{\sin \theta}.\label{eq:Chebyshev}
%\ee
$u=(m_{11}+m_{22})/2$:
\be
U_N(u) = \frac{\sin[(N+1)\cos^{-1}u]}{\sqrt{1-u^2}}.\label{eq:Chebyshev}
\ee
Using this rigorous result, we derive an exact form of the transfer matrix:
\be
M =
%\begin{pmatrix}
%(u+\alpha)U_{N-1}-U_{N-2} & (v-\beta)U_{N-1} \\
%(v+\beta)U_{N-1} & (u-\alpha)U_{N-1} -U_{N-2}
%\end{pmatrix}
\begin{pmatrix}
u+\alpha & v-\beta \\
v+\beta & u-\alpha
\end{pmatrix}U_{N-1}(u) - \mathbb{1}U_{N-2}(u), \label{eq:M_analytical}
\ee
where we have defined $v \equiv (m_{11}-m_{22})/2$, $\alpha \equiv (n_0m_{12}+m_{21}/n_0)/2$, and $\beta \equiv (n_0m_{12}-m_{21}/n_0)/2$. One can check that the elements of $M$ satisfy the same relations (\ref{eq:symm}) with $m_{pq}$ replaced by $M_{pq}$ and $\det{M}=(\det{M_j})^N=1$.
%from which the CPA-laser condition $M_{11}=M_{22}=0$ takes the following form:
%\begin{gather}
%(m_{11}+m_{22})U_{N-1} = U_{N-2}\\
%(m_{21}+m_{12}n_0^2)U_{N-1}=0.
%\end{gather}
%\begin{gather}
%2\left[|c|^2-(n_r^2-n_i^2)\left|\frac{s}{n}\right|^2\right]U_{N-1} = U_{N-2}\label{eq:Sol1}\\
%\left[sc^*(n+\frac{n_0^2}{n})+c.c.\right]U_{N-1}=0.\label{eq:Sol2}
%\end{gather}

The corresponding scattering matrix is then given by
\be
S =
\frac{
\begin{bmatrix}
(v-\beta)U_{N-1}(u) & 1\\
1 & -(v+\beta)U_{N-1}(u)
\end{bmatrix}
}
{(u-\alpha)U_{N-1}(u)-U_{N-2}(u)}.\label{eq:S_analytical}
\ee
Using this exact form of the scattering matrix, we again probe the spectral signatures of the CPA-laser point at the boundary of the Brillouin zone. A CPA-laser point takes place when a pole and a zero of $S$ becomes real and identical (see the discussions in the Appendix), and the pole condition is satisfied when the denominator in Eq.~(\ref{eq:S_analytical}) becomes zero, or equivalently,
\be
\frac{\sin\theta}{\tan N\theta} = \alpha,
\ee
where $\cos\theta\equiv u$. To the leading order of $n_i$ and at the boundary of the Brillouin zone, we find $\alpha\approx0$ and $\sin\theta\approx2n_i/n_r\approx\theta$, which leads to $\cos N\theta\approx\cos(2Nn_i/n_r)\approx0$. This result is equivalent to  Eq.~(\ref{eq:n_i}) obtained from the coupled mode theory.

\begin{figure}[t]
\centering
\includegraphics[width=\linewidth]{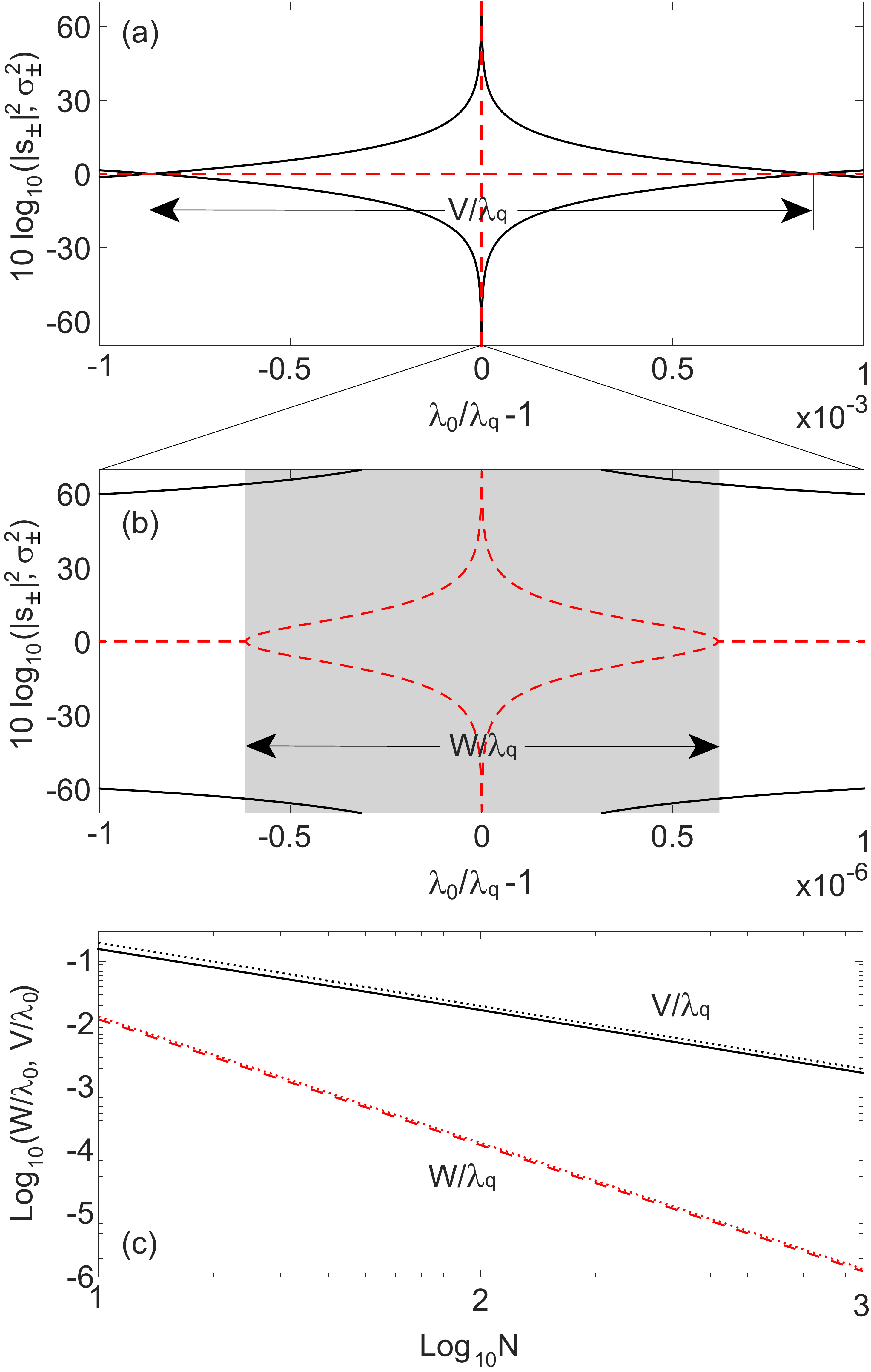}
\caption{(Color online) Contrasting spectral signatures of a CPA-laser calculated using the exact scattering matrix. (a) Same as Fig.~\ref{fig:CMT}. (b) Zoomed view of (a) near the CPA-laser point. Shaded area indicates the $\cal PT$-broken phase of the scattering matrix.
(c) $W$ and $V$ as a function of system size $N$. Numerical results of Eq.~(\ref{eq:S_analytical}) are shown by the solid and dashed lines. The perturbation results (\ref{eq:V}) and (\ref{eq:W}) are shown by the dotted lines. The value of $n_i$ is $N$-dependent and chosen according to Eq.~(\ref{eq:n_i}) with $P=1$.} \label{fig:exact}
\end{figure}

Similar to the quantitative knowledge we gained from the coupled mode theory, we find that the CPA-laser has a more prominent spectral feature in the singular value spectrum of $S$ than its eigenvalue spectrum [see Figs.~\ref{fig:exact}(a) and \ref{fig:exact}(b)]. It is striking to see that $V^{-1}$ scales linearly with the system size $N$ while $W^{-1}$ scales quadratically with $N$ (see the numerical results shown in Fig.~\ref{fig:exact}(c); solid and dashed lines). For example, $V$ is roughly 10 times $W$ when $N=10$, and this ratio increases dramatically to 1000 when there are 1000 periods. If we take the wavelength to be near infrared in the latter case, then $V$ is on the order of nanometers while $W$ is on the order of picometers.

To gain some analytical insights of these results, we expand $m_{pq}$ to the leading order of $n_i$, which gives $u \approx c_0^2 - s_0^2, \, v=2in_is_0^2/n_r, \,\alpha \approx 2is_0c_0, \, \beta=0$, where $c_0,s_0$ are defined similarly to $s,c$ but with $n_j$ replaced by $n_r$. By substituting these values into Eq.~(\ref{eq:S_analytical}), we find the singular values of $S$:
%$S=M_{22}^{-1}(\begin{smallmatrix}M_{12} & 1\\1 & -M_{21}\end{smallmatrix})$
\be
\sigma_\pm^2 = \frac{[1\pm vU_{N-1}(u)]^2}{|(u-\alpha)U_{N-1}(u)-U_{N-2}(u)|^2}.
\ee
They become equal when $U_{N-1}(u)=0$, from which we derive
\be
\frac{V}{\lambda_0} \approx 2N^{-1}.\label{eq:V}
\ee
Note that $V$ still marks the distance between two input-insensitive points, a property inherent to the singular value spectrum and independent of the exact form of the scattering matrix.
Similarly, by keeping the terms proportional to $n_i^2$, we find
\be
\frac{W}{\lambda_0}\approx\frac{\pi}{4}\sqrt{\frac{\pi^2}{2}-2}\,N^{-2}, \label{eq:W}
\ee
which again marks the size of the $\cal PT$-broken phase of $S$. These scaling behaviors agree well with the numerical results we have seen in Fig.~\ref{fig:exact}(c).

The contrasting values of $V$ and $W$ indicate that near perfect absorption and extremely strong amplification can also be achieved in the $\cal PT$-symmetric phase of the scattering matrix (see Fig.~\ref{fig:exact}(b), for example), which is known for and defined by its flux-conserving eigenstates as we have mentioned. This feature is a manifestation of the strong non-Hermicity of the system, %: for a Hermitian or unitary operator, its eigenstates form a complete and orthogonal basis, and the squares of its singular values are the same as those of the absolute values of its eigenvalues; in a non-Hermitian system, its eigenstates are no longer orthogonal, 
where the squares of its singular values are no longer bounded by those of the absolute values of its eigenvalues.

\section{Sensitivities of a CPA-laser}

Interestingly, the maximum and the minimum of the extinction ratio given by $\sigma_\pm^2$ do agree with $|s_\pm|^2$ in the original CPA experiment \cite{CPA_exp}, even though the system (a uniform silicon slab) is also non-Hermitian (and $S$ non-unitary) due to the presence of loss. This coincidence is a result of the mirror symmetry of the system \textit{and} optical reciprocity: $S$ contains identical diagonal elements (i.e., equal reflection coefficients from the two sides) due to the mirror symmetry and identical off-diagonal elements (i.e., the transmission coefficient) due to optical reciprocity. Consequently, $S^\dagger$ has the same eigenvectors as $S$ itself, which are given simply by symmetric and antisymmetric inputs of equal amplitudes. $\sigma_\pm^2$, given by the eigenvalues of $S^\dagger S$, are then equal to $|s_\pm|^2$.

By definition, a $\cal PT$-symmetric system does not possess mirror symmetry if there is gain and loss. Therefore, we do not expect the extinction ratio $\Theta$ in Eq.~(\ref{eq:ratio}) to be bounded by $|s_\pm|^2$. However, if we relax the requirement of a $\cal PT$-symmetric periodic structure, i.e., allowing it to have an additional layer of gain or loss, then the system does have mirror symmetry, with which $\sigma_\pm^2=|s_\pm|^2$ holds. Note that although we have mentioned in the introduction that CPA-lasing is a property of true $\cal PT$-symmetric systems, there are exceptions in other types of systems, such as when accidental degeneracies of a pole and a zero of the scattering matrix  take place \cite{Mostafazadeh} or when an imaginary permittivity and permeability satisfying $\varepsilon=-\mu$ are engineered \cite{EM}. Here this periodic structure with one additional layer of gain or loss is another exception: the CPA-laser point persists at the boundary of the Brillouin zone, which has the following simple physical interpretation. Since the effective wavelength inside the periodic structure is twice the lattice constant, the state with intensity peaks only in the loss layers gives rise to CPA and the state with intensities peaks only in the gain layers leads to lasing \cite{CPALaser_exp}. The additional layer of gain and loss does not break these patterns, and hence a CPA-laser point remains at the same wavelength.

\begin{figure}[b]
\centering
\includegraphics[width=\linewidth]{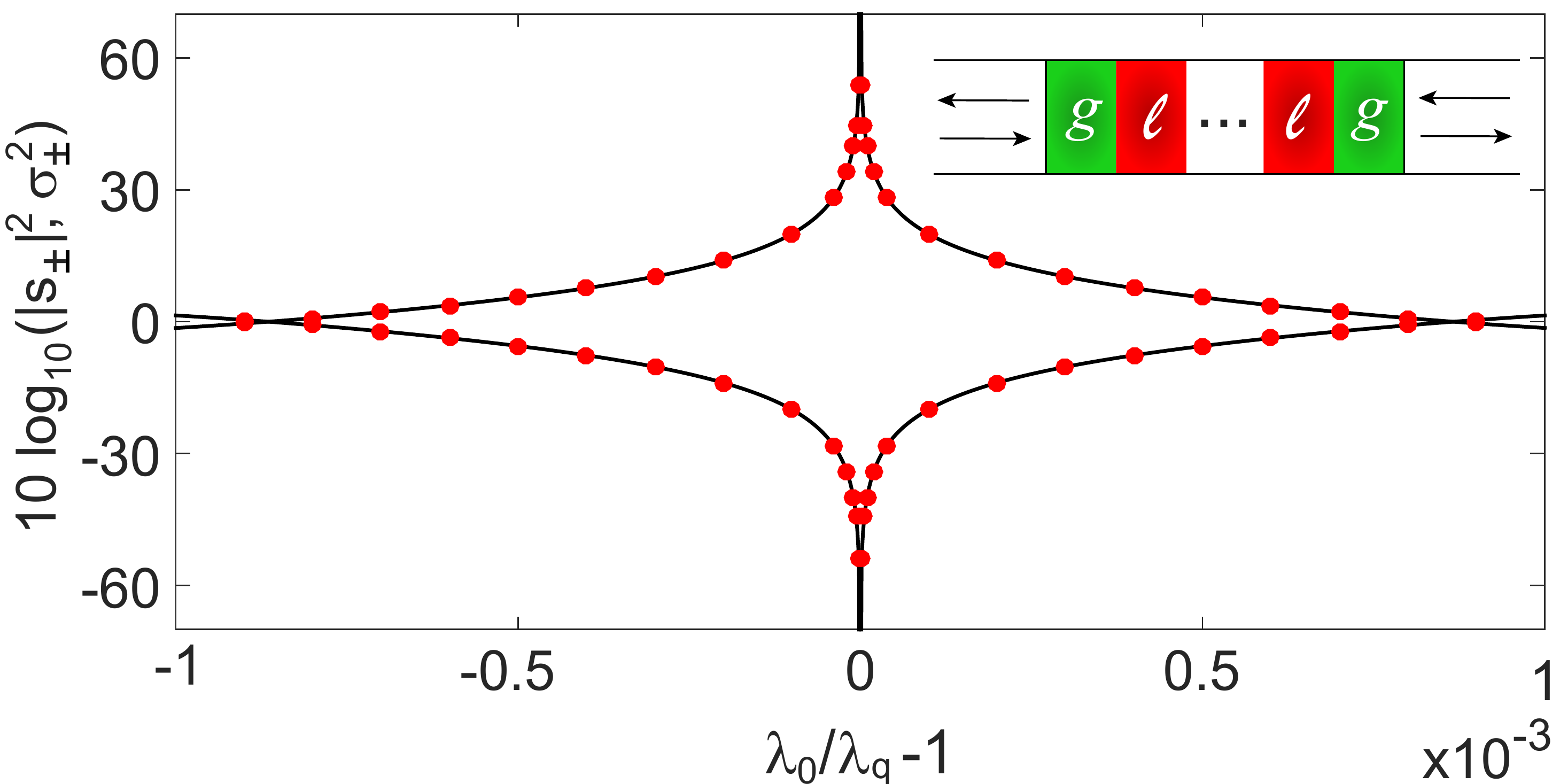}
\caption{(Color online) Identical spectral signatures of a symmetric CPA-laser. The solid lines show the singular value spectrum and dots show the eigenvalue spectrum. Inset: The structure has one additional layer of gain added to the $\cal PT$-symmetric structure in Figs.~\ref{fig:exact}(a) and (b), i.e., with 1001 layers of gain and 1000 layers of loss.
%The case with one additional layer of loss has the same spectra by considering the time-reversed processes.
} \label{fig:nonPT}
\end{figure}

In Fig.~\ref{fig:nonPT} we show that the spectral signatures $W$ and $V$ now become identical as expected. We note that the additional layer barely changes the singular value spectrum but completely alters the eigenvalue spectrum, which suggests that the latter is a good indicator of the true $\cal PT$ symmetry of the system.
In fact, it is likely that the waveguide structure used in Ref.~\cite{CPALaser_exp} was similar to this symmetric one here: the gain in the experiment was provided by the quantum wells on the chip and excited using a relatively long pump spot. Therefore, there were two sections of the waveguide outside the intended $\cal PT$-symmetric region, one on each end, where gain was activated. Nevertheless, we have checked that the spectral signatures of the CPA-laser vary little even when the lengths of these two gain sections are several times the lattice constant, which is an indication of the robustness of this period system for CPA-laser operation.

\begin{figure}[t]
\centering
\includegraphics[width=\linewidth]{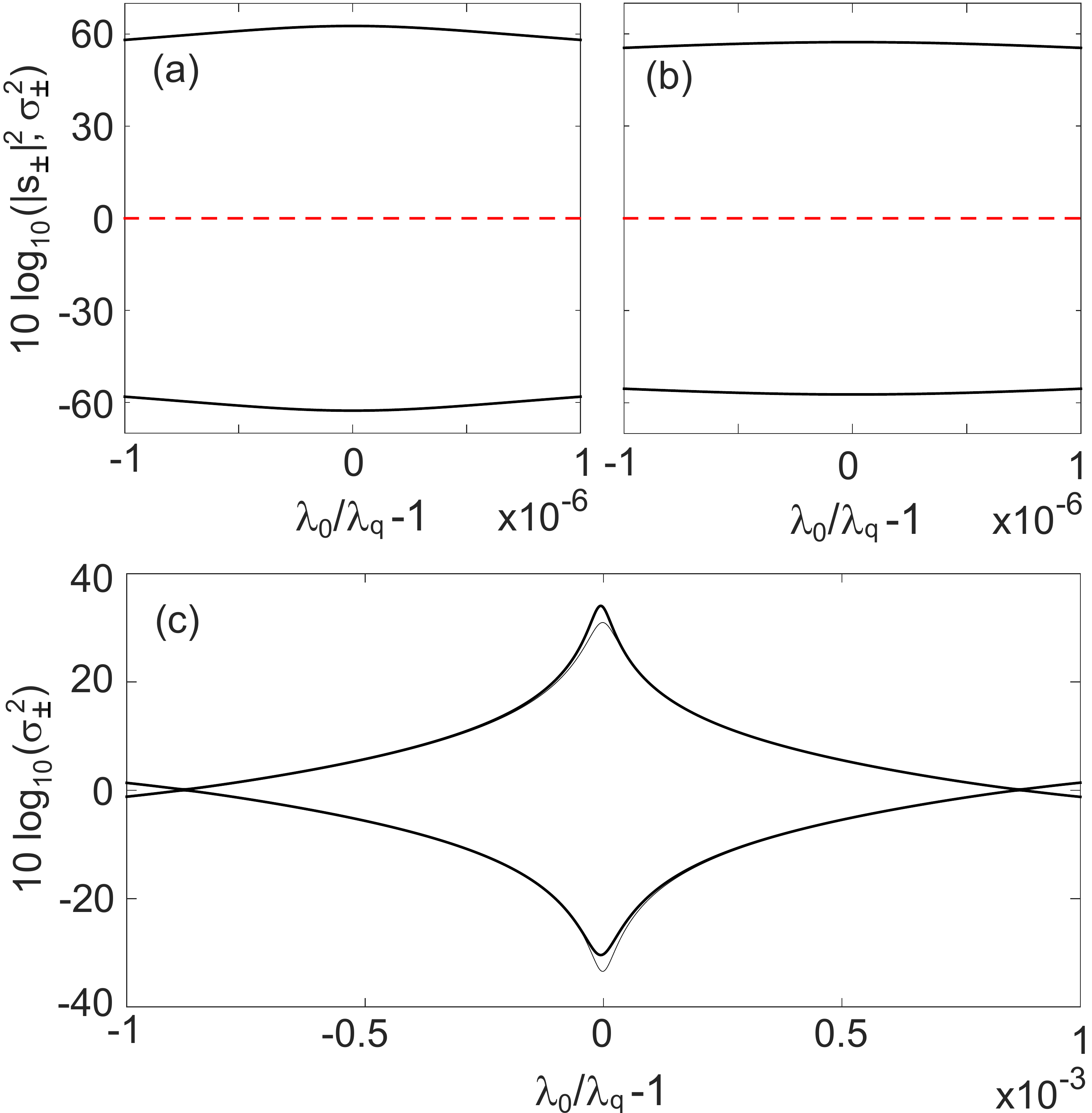}
\caption{(Color online) Sensitivities of CPA-laser to index variations and system disorder. In (a) $n_i$ differs from its value used in Fig.~\ref{fig:exact}(a) by $2.4\times10^{-6}$. In (b) $n_r$ differs from its value used in Fig.~\ref{fig:exact}(a) by $3\times10^{-3}$. In (c) both $n_r$ and $n_i$ are subjected to a Gaussian disorder with a standard deviation 1/1000 their values in Fig.~\ref{fig:exact}(a), and the length of each layer also follows a Gaussian distribution with a standard deviation of $a/25$, which is about $10$ nm when the CPA-laser wavelength $\lambda_q$ is near 1.5 ${\mu}m$. Two examples (thick and thin lines) are shown in (c).} \label{fig:sensitivity}
\end{figure}

The extremely sensitivity of the eigenvalue spectrum near the CPA-laser point can also be revealed by a slight variation of the refractive index, especially in its imaginary part. For the system with $N=1000$ shown in Figs.~\ref{fig:exact}(a) and (b), we find that a change of $n_i$ on the order of $10^{-6}$ from the ideal value given by Eq.~(\ref{eq:n_i}) is enough to completely eliminate the diverging feature of the eigenvalue spectrum at the CPA-laser point [see Fig.~\ref{fig:sensitivity}(a)], whereas the same deed requires a change of $n_r$ on the order of $10^{-3}$ [see Fig.~\ref{fig:sensitivity}(b)]. This difference can be understood from Eq.~(\ref{eq:n_i}): to keep the system at the CPA-laser point, a small variation of $n_i$ would require a much larger compensation from the change of $n_r$ when the system size is large. This sensitivity of the eigenvalue spectrum may be potentially utilized to detect environmental changes via $n_i$ if the periodic system is clean and $\cal PT$-symmetric.

Although the singular value spectrum is also affected when the index varies, the output intensity contrast between the absorbing and amplifying states is still outstanding as can be seen from Figs.~\ref{fig:sensitivity}(a) and (b), which indicates the robustness of CPA-laser operation in such a system. It can be shown that right at the boundary of the Brillouin zone, absorption and amplification disappear only when $n_i$ equals the value given by Eq.~(\ref{eq:n_i}) with $P$ being a half integer, including the trivial case $n_i=0$. To reach these points $n_i$ needs to vary from its value at the CPA-laser point by at least 100\%, and hence a small fractional change of $n_i$ retains a large contrast between the extinction ratios in the absorbing and amplifying state. 

This robustness holds even after taking the disorder of the system into consideration. For example, in Fig.~\ref{fig:sensitivity}(c) we introduce a
Gaussian disorder to both the length and index of each layer. Although the highest gain and loss is reduced from the ideal case shown in Fig.~\ref{fig:exact}(a), an intensity contrast of more than 60~dB still exists between the output light in the amplifying state and the absorbing state. Therefore, even though $\cal PT$ symmetry is destroyed by these disorders, the CPA-laser continues to perform well, similar to the case we have discussed with one additional layer of gain or loss.

So far we have discussed the contrasting sensitivities of the eigenvalue and singular value spectra of $S$ near the CPA-laser point when the system is subjected to imperfection or disorder. In what follows we turn to a different type of imperfection, which lies in the amplitude and phase mismatch of the input light from the CPA-laser condition. For simplicity, we first turn to the clean periodic structure considered in Fig.~\ref{fig:exact} that has $\cal PT$ symmetry. $S$ in a 1D $\cal PT$-symmetric system can be parametrized in general by \cite{conservation}
\be
S = \frac{1}{a}
\begin{pmatrix}
ib & 1\\ 1& ic
\end{pmatrix},\label{eq:S_para}
\ee
where $b,c$ are two real quantities and satisfy $bc = |a|^2-1$. The complex amplitude ratios of the two input beams in its two eigenstates are given by \cite{conservation}
\be
\varpi_\pm = \frac{i}{2}[(c-b)\pm\sqrt{(b-c)^2-4}].
\ee

In the $\cal PT$-broken phase $|b-c|>2$, and $\varpi_\pm$ are imaginary with the same sign. This is an interesting observation because it indicates that the phase difference between the two input beams is exactly $\pi/2$ in \textit{both} eigenstates of $S$, even though one of them gives rise to the CPA state and the other one is associated with the lasing state at the CPA-laser point. More importantly, this observation also indicates that away from the CPA-laser point but still in the $\cal PT$-broken phase, one can switch between a strongly absorbing state and a strongly amplifying state by simply changing the intensity ratio of the two input beams, which correspond to the two eigenstates of $S$. 

\begin{figure}[b]
\centering
\includegraphics[width=\linewidth]{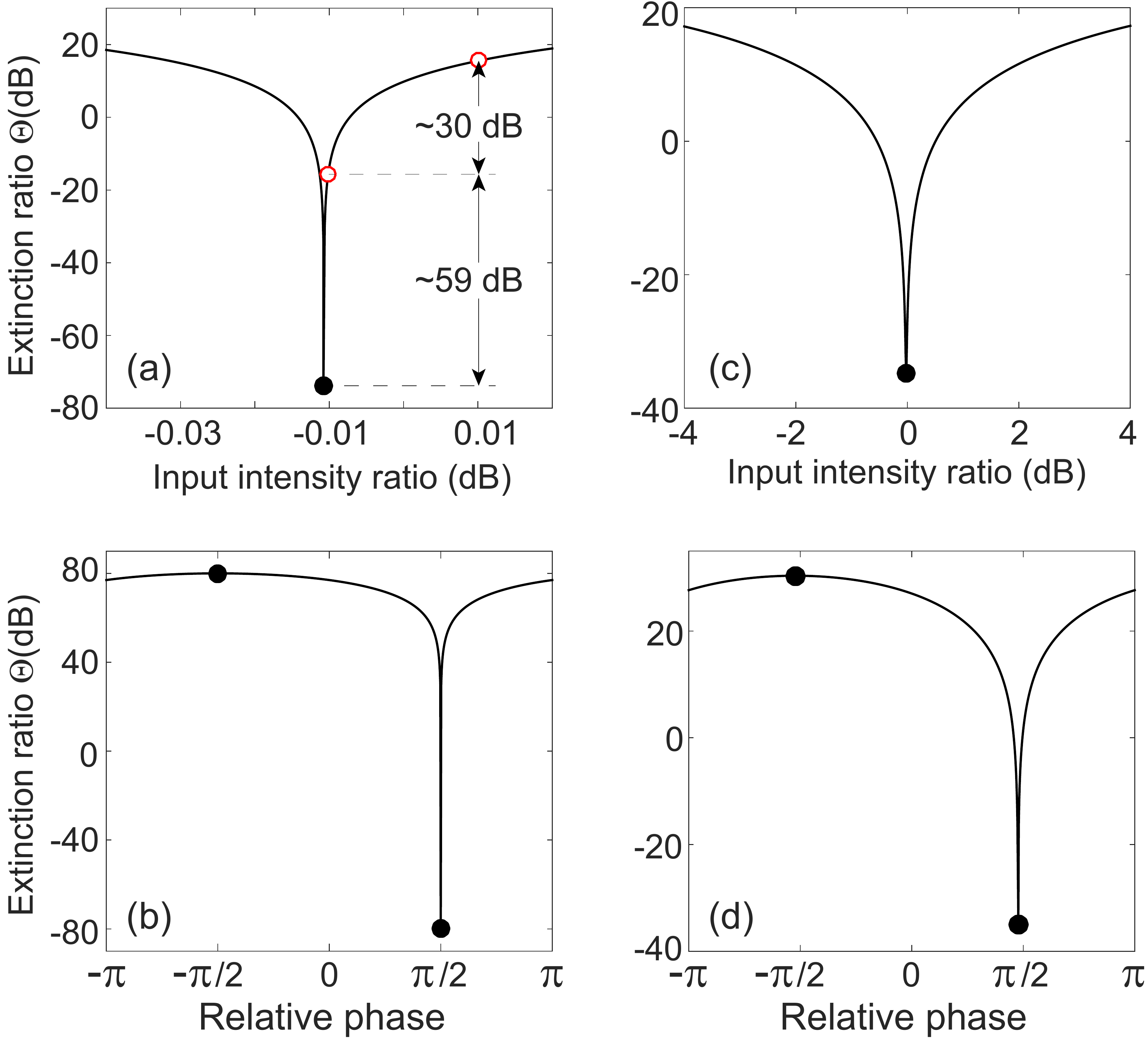}
\caption{(Color online) Sensitivities of a CPA-laser to the relative amplitude (a) and phase (b) of the two input beams. In (a) the relative phase of the two beams are fixed at $\pi/2$, with the open circles marking the scattering eigenstates of $S$ and the filled circle marking the right singular vector of $S$ in the maximally absorbing state. In (b) the relative amplitude of the two input beams is fixed at 1, 
%its value in the maximally absorbing state, 
with the filled circles marking the maximally absorbing and amplifying states. In both (a) and (b) the wavelength is slightly away from the CPA-laser condition as mentioned in the main text, and the clean system is the same as in Figs.~\ref{fig:exact}(a) and \ref{fig:exact}(b). (c) and (d) Similar to (a) and (b) but with one realization of the disorder considered in Fig.~\ref{fig:sensitivity}(c). In (c) the relative phase of the two input beams is fixed at its value in the maximally absorbing state, which is shifted away from $\pi/2$ due to disorder [see (d)]. %The eigenvalues of $S$ are not shown in (c) because the relative phases between the two input beams in them are no longer identical.
} \label{fig:phase}
\end{figure}

For example, in the $\cal PT$-symmetric system with $N=1000$ considered in Figs.~\ref{fig:exact}(a) and (b) and at the wavelength $2\times10^{-7}\lambda_0$ away from the CPA-laser point, changing the intensity ratio of the two input beams by merely 0.02 dB switches the system from the amplifying eigenstate to the absorbing eigenstate of $S$, resulting in a change of the extinction ratio $\Theta$ by more than 30 dB [see Fig.~\ref{fig:phase}(a)]. This feature may potentially be exploited as a differential amplifier of light.

Moreover, since the eigenvalues of $S$ are not the limits of $\Theta$, we can further increase the intensity contrast of the switching. For a quantitative analysis, we turn to the relative phase and amplitude of the right singular vectors of $S$, which correspond to the maximally amplifying and absorbing states. These right singular vectors are also the eigenvectors of $S^\dagger S$, which can be expressed as follows using Eq.~(\ref{eq:S_para}):
\be
S^\dagger S =
\frac{1}{|a|^2}
\begin{bmatrix}
1+b^2 & -i(b-c)\\ i(b-c) 1& 1+c^2
\end{bmatrix}.\label{eq:SS}
\ee
In our system $b\approx-i(v-\beta)U_{M-1}\approx-c\ll1$, then it is straightforward to see that the eigenstates of $S^\dagger S$ are $(i~~1)^T$ and $(-i~~1)^T$ to a very good approximation, with the former corresponding to the maximally absorbing state and ``$T$" denoting the matrix transpose. Therefore, by fixing the relative phase between the left and right input beam at $\pi/2$, we find that $\Theta$ in the absorbing state can be further reduced to $-74$ dB in Fig.~\ref{fig:phase}(a) when reducing the input intensity ratio by $10^{-3}$ dB from the absorbing eigenstate of $S$. The maximally absorbing state also shows good sensitivity to the relative phase of the input beams as we show in Fig.~\ref{fig:phase}(b), while its amplifying partner does not.

These extraordinary sensitivities can be potentially utilized to detect a minute amplitude and phase change of a coherent signal. For example, one can imagine splitting a coherent signal from an external laser source into two beams of similar amplitudes and feeding them into the two sides of a CPA-laser. In this setup one beam after the splitter can be exposed to a region where there are events that change the phase or amplitude of this beam. These events can then be detected by observing $\Theta$ at the output ports of the CPA-laser. Unlike a standard interferometer, this setup can also detect a minute amplitude difference occurred when light propagates along the two legs.

From the discussions above, we see that the amplitude sensitivity can be attributed to the eigenvalue spectrum of $S$, whereas the phase sensitivity can be associated with the singular value spectrum of $S$. These sensitivities are reduced in the presence of disorder and imperfection that destroy the $\cal PT$ symmetry of the system [see Figs.~\ref{fig:phase}(c) and (d)]. While detrimental for sensing applications, they reduce the challenge to tune into the maximally absorbing state and hence increases the feasibility of building and operating a CPA-laser.

\section{Conclusion and discussion}

Using a coupled mode theory and a rigorous transfer matrix calculation, we have shown that a CPA-laser in a $\cal PT$-symmetric periodic structure can display contrasting spectral signatures in the eigenvalue spectrum and the singular value spectrum of the scattering matrix. These two spectra are of particular interest because the eigenvalue spectrum defines the $\cal PT$ symmetry breaking of a scattering system, while the singular value spectrum imposes the limits of absorption and amplification in a CPA-laser under realistic conditions. The quantitative measures of these two spectra, $W$ and $V$, scale with the system size linearly and quadratically, which lead to their vastly different values when the system size is large. Another length-related behavior in $\cal PT$-symmetric periodic systems is the disappearance of unidirectional invisibility \cite{Lin,conservation,Feng2}, where perfect transmission and vanished reflection from one side of the system take place only when the system size is small \cite{Longhi_PTBragg}.

We have also discussed the contrasting sensitivities of these two spectra for different types of imperfection and disorder, for both the system itself and for the two coherent input beams. On the one hand, thanks to the same relative phase between the latter in the amplifying and absorbing eigenstates of $S$, we have suggested a differential amplifier for light, which may also be incorporated into an interferometric setup to detect minute phase and amplitude differences along its two legs. On the other hand, the singular value spectrum attests to the robust of a CPA-laser in a periodic structure, even when the $\cal PT$ symmetry does not hold due to disorder and experimental implementation. Note that we have not included quantum or thermal noises in our analysis, which will impose additional limits on the sensitivity of using a CPA-laser as a detector \cite{noise}.
We also remark that %while a CPA-laser is associated with the vanishing and diverging eigenvalues of $S$, probing the lasing state  requires a different relative phase (i.e., $-\pi/2$) between the two input beams from that in the corresponding eigenstate of $S$ (i.e., $\pi/2$). There is no such issue for probing the CPA state, where this relative phase is the same as that in the corresponding eigenstate of $S$ (also $\pi/2$). Also
if one attempts to use the other definition of $S$, i.e., with the transmission coefficient on the diagonal (see the discussions in Ref.~\cite{conservation}), one losses the sensitive information contained in spontaneous $\cal PT$ symmetry breaking, because its $|s_\pm|^2$ are almost identical to $\sigma_\pm$ here (not shown).

Finally, we have argued in Sec.~\ref{sec:CMT} that $|\sigma_+\sigma_-|=1$ using the $\cal PT$-symmetry of the system. Now utilizing the general expression (\ref{eq:SS}) for $S^\dagger S$ and the property $bc=|a|^2-1$ \cite{conservation}, we find $\sigma_+\sigma_-=(1+bc)^2/|a|^2=1$, which further specifies the opposite signs of $\sigma_\pm$. Although this property is similar to that of the eigenvalues of $S$ in the $\cal PT$-broken phase (i.e., $s_+s_-^*=1$), the singular value spectrum of $S$ does not undergo a spontaneous symmetry breaking. This is because the left (right) singular vectors of $S$ are also the eigenstates of the Hermitian matrix $SS^\dagger$ ($S^\dagger S$), and as a result, these vectors form an orthogonal basis and cannot coalesce. We also know from basic matrix theory that singular values are non-negative real numbers, and hence they cannot become complex conjugate pairs.

\section*{acknowledgement}

L.G. acknowledges support by NSF under grant No. DMR-1506987. L.F. acknowledges support by NSF under grant DMR-1506884.

\appendix
\section*{Appendix: Poles and zeros of the $S$ matrix}
\label{sec:app}

In the main text we treated $k_0$ as a real quantity when deriving the exact form of the $S$ matrix. If a complex $k_0$ is considered, then Eq.~(\ref{eq:M_period}) is modified to
\be
M_j =
\begin{bmatrix}
c\tilde{c} - \frac{n^*s\tilde{s}}{n} & i\frac{s\tilde{c}}{n}+i\frac{c\tilde{s}}{n^*} \\
i{n}{s\tilde{c}}+i{n^*}{c\tilde{s}}  & c\tilde{c}- \frac{ns\tilde{s}}{n^*}
\end{bmatrix},
\ee
where $\tilde{s}\equiv\sin(n_j^*k_0a/2), \tilde{c}\equiv\cos(n_j^*k_0a/2)$. $M_j$ satisfies the constraints imposed by $\cal PT$ symmetry \cite{Longhi}:
\be
m_{11}(k_0) = m_{22}^*(k_0^*),\quad m_{12,21}(k_0) = -m_{12,21}^*(k_0^*).\label{eq:symm2}
\ee
The analytical form of the $S$ matrix given by Eq.~(\ref{eq:S_analytical}) still holds with the modified $M_j$, and one can show that a pole and zero of $S$ form a complex conjugate pair \cite{CPALaser}.

\begin{figure}[bth]
\centering
\includegraphics[width=\linewidth]{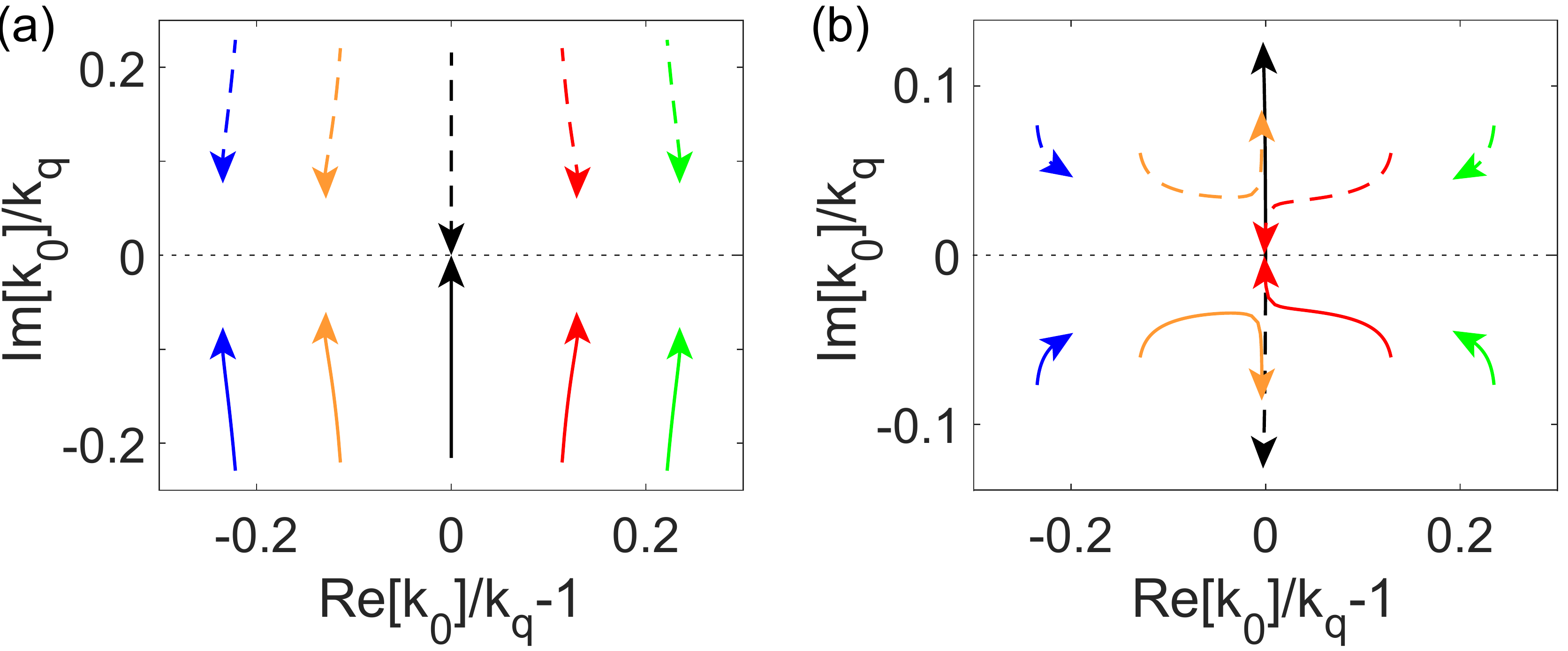}
\caption{(Color online) Trajectories of $S$ matrix poles (solid) and zeros (dashed) when $n_i$ increases. In (a) $n_i\in[2.5\times10^{-3},~0.25]$ and the central pair of pole and zero meet on the real axis. In (b) $n_i\in[0.25,~0.72]$ and the second pair from the right meet on the real axis. The $\cal PT$-symmetric structure is chosen to be shorter ($N=10$) than that in Fig.~\ref{fig:exact}(a), with all other parameters kept the same. $k_q\equiv2\pi/\lambda_q=q/2n_r$ is the free-space wave number at the boundary of the Brillouin zone.} \label{fig:poleZero}
\end{figure}

Note that since the $\cal PT$-symmetric structure we considered is differentiated from the leads only by the weak modulation of the imaginary part of its refractive index, there is no resonance or anti-resonance when $n_i=0$. As $n_i$ becomes nonzero and increases, there is one pair of pole and zero emerging from $\im{k_0}=\pm\infty$ that move almost vertically toward each other [see Fig.~\ref{fig:poleZero}(a)]. This behavior is qualitatively different from the case where gain and loss is added gradually to an existing cavity (i.e., $n_r\neq n_0$), where two poles (zeros) undergo an anti-crossing with only one of them leading to a CPA-laser point \cite{CPALaser}.

In the current case, this pair of pole and zero eventually meet on the real axis when the CPA-laser condition is first satisfied, i.e., when $n_i$ equals the value given Eq.~(\ref{eq:n_i}) with $P=1$. Afterwards they continue to move into the other vertical half of the complex $k_0$ plane. Note that at this CPA-laser condition with $P=1$, there are other poles (zeros) with a finite imaginary part. They undergo anti-crossing pairwise as $n_i$ further increases, just like in the $n_r\neq n_0$ case; each pair of poles (zeros) also only gives rise to a single CPA-laser point [see Fig.~\ref{fig:poleZero}(b)], at the value of $n_i$ given by Eq.~(\ref{eq:n_i}) with $P>1$. In Fig.~\ref{fig:poleZero} the periodic structure is relatively short and the values of $n_i$ at the CPA-laser points deviate slightly from Eq.~(\ref{eq:n_i}).

\bibliographystyle{longbibliography}

\end{document}